\documentclass[pra,twocolumn,english,showpacs,floatfix,superscriptaddress]{revtex4-1}

\usepackage{amsmath}
\usepackage{amssymb}

\usepackage{amsmath,amssymb}
\usepackage[usenames]{color}
\usepackage{mathbbol}
\usepackage{amssymb}
\usepackage{grffile}
\usepackage[pdftex]{graphicx}
\usepackage{amsmath, amstext, amssymb, amsfonts, amsxtra}
\usepackage{textcomp}
\usepackage{xspace} 
\usepackage[colorlinks]{hyperref}

\DeclareSymbolFontAlphabet{\amsmathbb}{AMSb}

\newcommand{\bonnpi}{Physikalisches Institut, University of Bonn, Nussallee 12, 53115 Bonn, Germany}
\newcommand{\optimas}{Department of Physics and research center OPTIMAS, University of Kaiserslautern, Germany}
\newcommand{\mainz}{Graduate School Materials Science in Mainz, Gottlieb-Daimler-Strasse 47, 67663 Kaiserslautern, Germany}
\newcommand{\innsbruck}{Institut f\"ur Experimentalphysik und Zentrum f\"ur Quantenphysik, Universit\"at Innsbruck, 6020 Innsbruck, Austria}

\begin{document}
\title{The Talbot effect in the presence of interactions}

\author{Philipp H\"ollmer}
\affiliation{\bonnpi}

\author{Jean-S\'ebastien Bernier}
\affiliation{\bonnpi}

\author{Corinna Kollath}
\affiliation{\bonnpi}

\author{Christian Baals}
\affiliation{\optimas}
\affiliation{\mainz}

\author{Bodhaditya Santra}
\affiliation{\optimas}
\affiliation{\innsbruck}

\author{Herwig Ott}
\affiliation{\optimas}

\begin{abstract}
  We study both experimentally and theoretically, considering bosonic atoms in a periodic potential, the influence of interactions in a Talbot interferometer. While interactions decrease the contrast of the revivals, we find that over a wide range of interactions the Talbot signal is still proportional to the phase coherence of the matter wave field. Our results confirm that Talbot interferometry can be a useful tool to study finite range phase correlations in an optical lattice even in the presence of interactions. The relative robustness of the Talbot signal is supported by the first demonstration of the three-dimensional Talbot effect.
\end{abstract}

\date{\today}
\maketitle

\section{Introduction}

The Talbot effect describes periodic revivals of the interference pattern behind a diffraction grating. It is one of the most prominent near-field interference phenomena and was first discovered in 1836 for light \cite{Talbot1836}. 
In the last decades, the Talbot effect and various variations have seen a revival in the field of matter wave interference. To date, it is the best suited approach to study interference of large molecules \cite{Brezger2002,Gerlich2007}, reaching record values of more than 10000 atomic mass units \cite{Eibenberger2013}. Ultracold atoms also offer a perfect platform to study the Talbot effect for matter waves \cite{Chapman1995,Deng1999,Mark2011}.

In ultracold quantum gases, global phase coherence is routinely studied in time of flight imaging~\cite{BraunSchneider2015,BlochZwerger2008}. Finite range phase coherence, however, is more challenging to access. It has been accomplished by coherently outcoupling atoms from a condensate and studying them by a high-finesse optical cavity~\cite{RitterEsslinger2007, DonnerEsslinger2007} and by considering samples in reduced dimensions, which interfere with a reference twin system~\cite{HofferberthSchiedmayer2007,GringSchmiedmayer2012,ChomazDalibard2015}. Some of the current authors have recently demonstrated that the Talbot effect can be used to study finite-range phase in optical lattices \cite{Santra2017}. 

In optics, the Talbot effect can be conveniently observed with standard diffraction gratings. Behind the grating, the intensity distribution of the light field periodically matches the diffraction grating pattern at multiples of the Talbot distance, $L_T=2d^2/\lambda$, where $d$ is the grid size and $\lambda$ is the wavelength. Being a near-field interference effect, the revivals are sensitive to the phase coherence between adjacent slits of the diffraction grating.
As the propagation path becomes longer, the spatial area from which individual light beams interfere within a particular revival gets wider.
Later revivals therefore probe larger range spatial phase coherence. The observation of a reduced contrast for later revivals is an indicator for vanishing phase coherence. The assumption of single particle interference in a Talbot interferometer is ideally fulfilled in light interferometers, but not necessarily justified for matter wave fields.

While in standard molecular beam experiments, the beam intensity can be reduced such that collisions play no role, this might not be the case in other realizations at higher number densities, as it is the case in ultracold atoms in optical lattices. Collisions and interactions between the particles during the interferometer sequence can occur. For example, atomic collision during free expansion are clearly visible in s-wave scattering halos of atomic clouds expanding from an optical lattice \cite{Greiner2001}.

Interaction effects were only briefly mentioned in Ref.\,\cite{Santra2017} and a detailed understanding or theoretical modeling of interaction effects on the Talbot signal was not given. This is, however, important for a quantitative evaluation of Talbot interferometers in the presence of interparticle interactions. Therefore, we here study experimentally and theoretically the influence of interactions in a Talbot interferometer.
In section \ref{sec:experiment}, we perform experiments showing the dependence of the Talbot signal and the contrast of the Talbot revivals for bosonic atoms confined to a three-dimensional optical lattice. The effective interaction strenghts in the initial state is varied via the lattice depths. A clear dependence on the interaction strength can be observed. 

While an effective mean-field picture can be handled via nonlinear equations such as the Gross-Pitaevskii equation, the inclusion of individual collisions require an advanced microsopic modeling. In section \ref{sec:theory}, we develop a theoretical model for a temporal Talbot interferometer, which is capable to treat the expansion in free space. The model is based on a discretization of the space by a Bose-Hubbard model with an additional potential representing the optical lattice. We discuss the dependence of the Talbot signal on the interaction strength treating the interaction before and during the expansion separately. Our findings support the interpretation that the Talbot signal highlights the presence of finite range phase coherence in the initial state. Most deviations from this interpretation arise due to the presence of interaction during the expansion.   We conclude in section \ref{sec:3d} with the demonstration of a three-dimensional Talbot effect for a Bose-Einstein condensate in an optical lattice. Thereby, the atoms simultaneously interfere in all three spatial dimensions at high atomic densities.

\section{Experiment}
\label{sec:experiment}
In the experiment, we use a cigar-shaped $^{87}$Rb Bose-Einstein condensate 
residing in an optical dipole trap, which is superimposed by a three-dimensional 
optical lattice. The oscillation frequencies in the dipole trap are 
$\omega=2\pi\times 13\,$s$^{-1}$ and $\omega_{\perp}=2\pi\times 170\,$s$^{-1}$ 
and the total atom number is approximately $50000$. The lattice spacing parallel 
to the long condensate axis is $d=547\,$nm. The lattice spacing in the two 
perpendicular directions is a factor of $\sqrt{2}$ smaller and amounts to 
$d_{\perp}=387\,$nm. Details on the experimental setup can be found, e.g., in 
Refs.\,\cite{Gericke2008,Mullers2018}. The experimental sequence and the 
appearance of a Talbot interferometer signal has been discussed in a previous 
work for a one-dimensional optical lattice \cite{Santra2017}. In brief, we start 
by loading the condensate into the optical lattice for different lattice heights 
within 210\,ms. After a wait time of 50 ms, we then switch off instantaneously 
one lattice axis (in this section) or all lattice axes (in section \ref{sec:3d}) 
and let the atoms expand and interfere freely for a given evolution time. We 
then suddently switch on the original lattice configuration. All switching 
processes are shorter than $3\mu s$. In the ideal situation of a non-interacting 
atom cloud, the matter wave field undergoes revivals during the free expansion, 
which occur at multiples of the Talbot time $T_T$ given by 

\begin{equation}
T_T=\frac{2md^2}{h}
\label{eq:Talbot_time}
\end{equation}
where $h$ is Planck's constant and $m$ the mass of the atoms. Without 
interactions, the revivals are perfect and the switching on process of the 
lattice after the free evolution projects back onto the original wave-function. 
In the presence of interactions, however, the matter wave field at multiples of 
the Talbot time can differ from the initial one and the sudden switching on of 
the lattice can excite the atomic ensemble. Therefore, we wait for $100\,$ms to 
allow the system to rearrange before we measure the atomic density distribution 
in a standard time-of-flight image \cite{Santra2017}.

\begin{figure}[h!]
\includegraphics[width=0.9\linewidth]{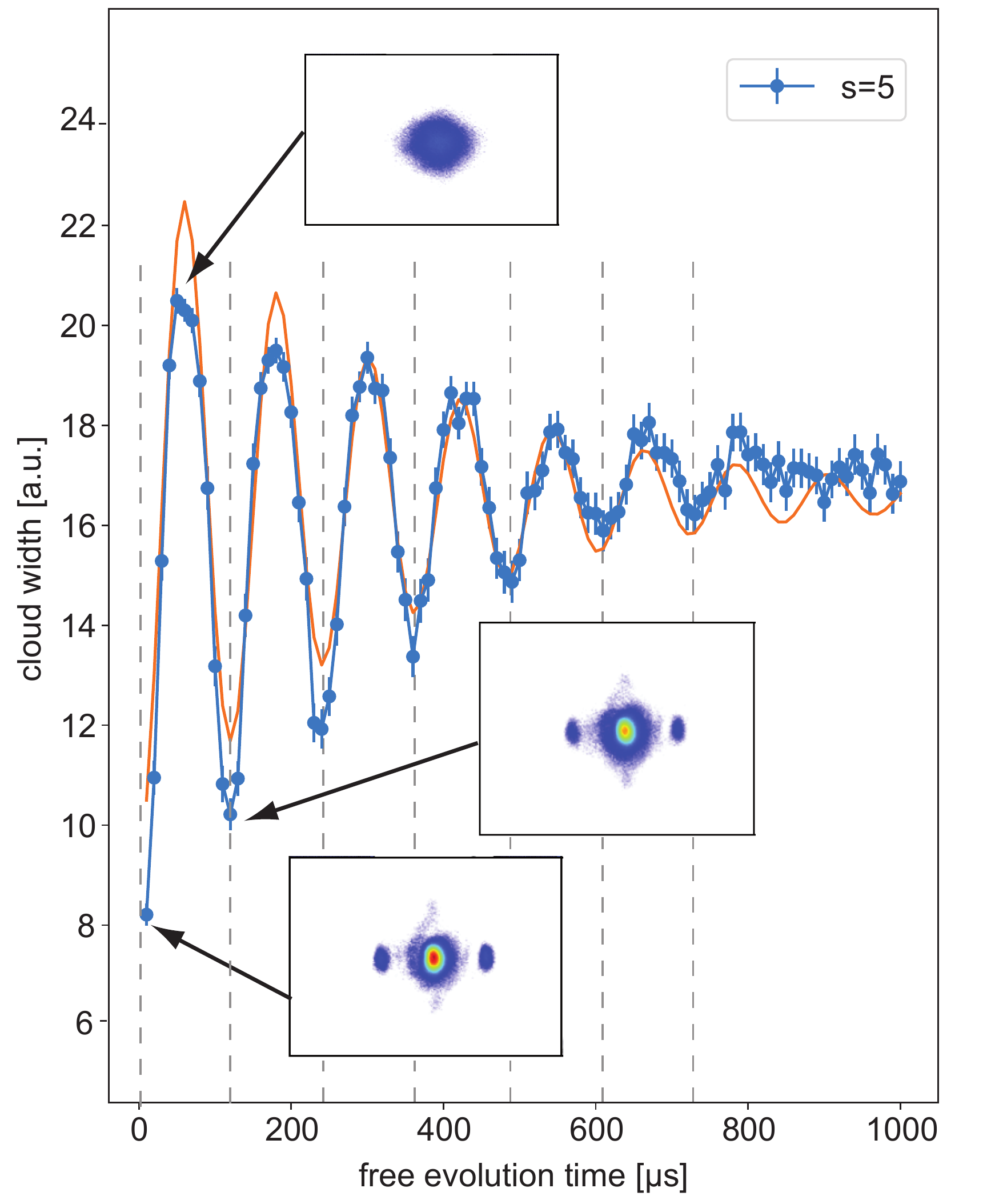}
\caption{Atomic Talbot interferometer in the presence of interactions. The plot shows the width of the central peak in the absorption image for different free evolution times between the switching off and on of one axis of a 3D optical lattice. The experimental data are fitted with an exponentially damped oscillation, from which we extract the decay time and the Talbot time ($T_T=120\,\mu$s). At multiples of the Talbot time, we observe periodic revivals of the matter wave interference pattern (dashed grey lines), while for times in between, the atomic cloud is heated up, leading to a smearing out of the interference pattern. The experiment was performed at a lattice potential of $V=s\times E_r$, with $s=5$, where $E_r=\hbar^2\pi^2/(2md^2)$ is the recoil energy of an atom with mass $m$ in a lattice with lattice constant $d$.}
\label{fig:experiment}
\end{figure}

In Fig.\,\ref{fig:experiment}, we show a typical Talbot signal for different evolution times. We plot the width of the atomic density distribution after the time-of-flight expansion, which is a direct measure of the excess energy in the system. The cloud size becomes minimal at mupltiples of $120 \mu s$. This is in good agreement with the Talbot time of $T_T\approx 130 \mu s$ calculated from the experimental parameters using Eq.~\ref{eq:Talbot_time} taking into account experimental imperfections as for example small misalignments of the laser beams.  

At the various multiples of the Talbot time, it is clearly visible that the amplitude of the width, the Talbot signal, decays as a function of time. For non-interacting atoms every individual revival, but also the anti-revivals, can be mapped to a phase correlator indicating the coherence, in the initial state,
between atoms located at a certain distance and confined to the optical 
lattice~\cite{Santra2017}. The observed decay then represents the decay of the 
phase correlations with distance. Thus, the strength of this kind of Talbot 
interferometer is its ability to measure finite range phase correlations in 
discrete lattice systems resolving the different distances. 

\begin{figure}[h!]
\includegraphics[width=1\linewidth]{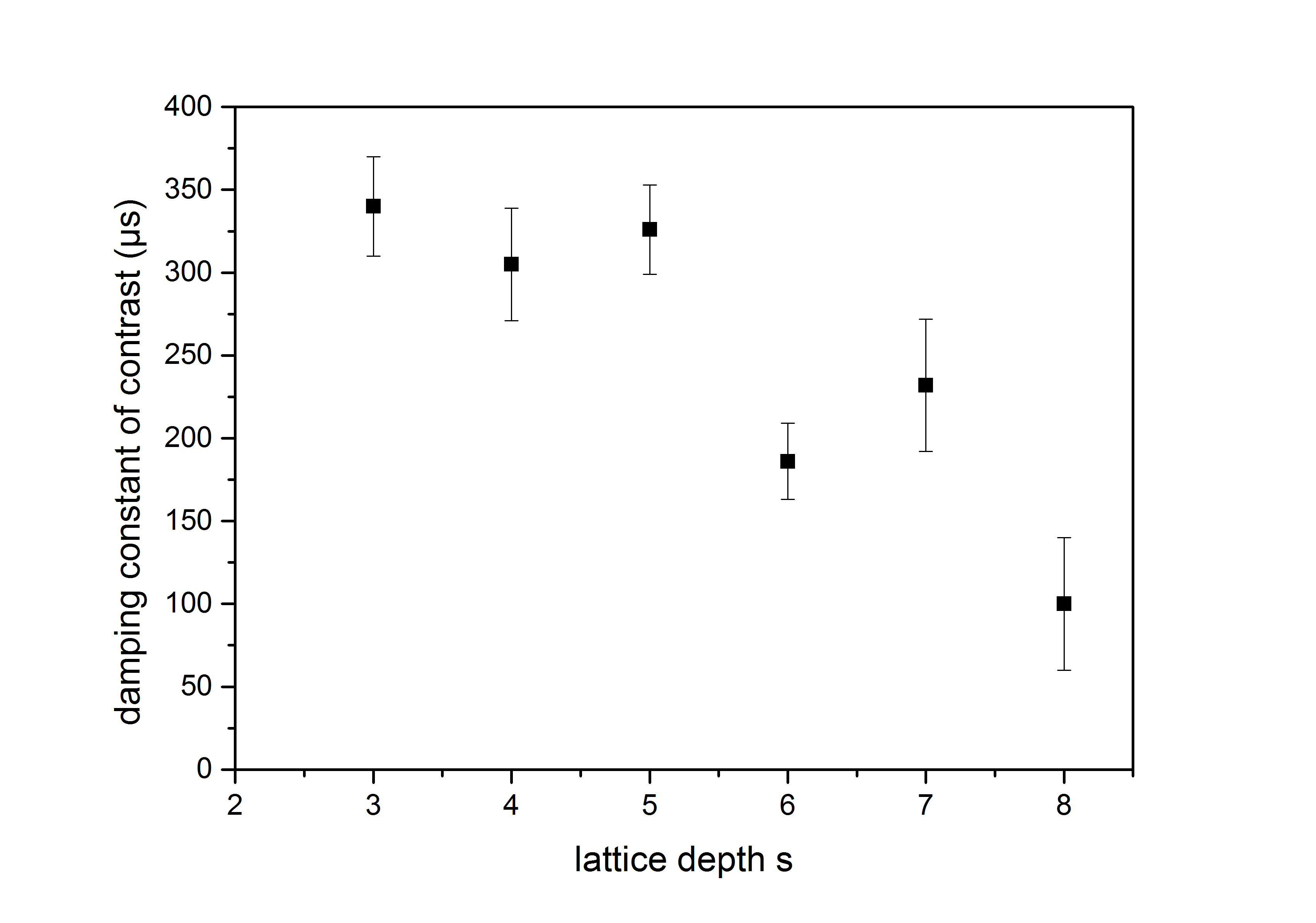}
\caption{Decay time of the Talbot signal as shown in Fig.\,\ref{fig:experiment} 
for different lattice heights. The signal decays faster for increasing optical 
lattice height corresponding to effectively stronger interactions between the 
atoms.}
\label{fig:damping}
\end{figure}

Here, we are interested in the influence of the interaction on the damping of 
the Talbot signal. Fig.\,\ref{fig:damping} shows the decay time of the Talbot 
signal for different lattice heights. The compression of the atoms in the 
lattice site increases the interaction energy in the optical lattice and the 
presence of the lattice potential decreases the kinetic energy, such that the 
ratio of the interaction to the kinetic energy is drastically changed. We find 
that the Talbot signal decays faster for deeper optical lattices corresponding 
to increasing effective interactions. Regarding the role of interactions, the 
reduced contrast can have two contributions. First, the initial state can 
already have reduced phase correlations due to the larger effective interatomic 
interactions. This is the interesting contribution to the signal, from which the 
finite range phase correlations could be extracted \cite{Santra2017}. Second, 
interactions during the free evolution, in particular, during the first moment 
of expansion from the lattice, could spoil the simple interference scheme of the 
Talbot effect and can lead to a further reduction of the contrast. It is 
important to understand this second contribution in order to discriminate 
between the two. In the experiment, however, we cannot change the interaction 
during the free expansion and a proper discrimination or a systematic 
investigation of the interaction effects is not possible.

\section{Theoretical results}
\label{sec:theory}
The qualitative results discussed above impose two important questions that we address in the theoretical modeling.
(i) How does the presence of interactions, both in the initial state and during the evolution, affect the Talbot signal?
(ii) Can the Talbot signal still be interpreted as a measurement of the finite size phase correlations of the initial state even in the presence of interactions?
We do not aim at a direct comparison with the experiment,
but rather want to give a first generic answer to these questions.

In the theoretical description, we assume the gas to be prepared in the ground state of the optical lattice and
focus on the time-evolution of the interacting gas after the sudden switch off of the lattice. For computational simplicity,
we consider a one-dimensional system, since we expect that the main effect of the influence of the interaction already
arises there, if not in a stronger fashion. The Hamiltonian
describing this one-dimensional system is (see e.g.~\cite{BlochZwerger2008})
\begin{align}
  H &= -\frac{\hbar^2}{2m} \int_0^L dx~\psi^\dagger(x) \frac{\partial^2}{\partial x^2} \psi(x) \nonumber \\
  &~~~ + \frac{g^\text{1D}(t)}{2} \int_0^L dx~\psi^\dagger(x) \psi^\dagger(x) \psi(x) \psi(x) \nonumber \\
  &~~~ + \int_0^L dx~V(x,t)~\psi^\dagger(x) \psi(x),
\end{align}
where $\psi(x)$ and $\psi^\dagger(x)$ are the
annihilation and creation operators for the $N$ bosonic atoms of mass $m$, $g^\text{1D}(t)$ is the repulsive contact
interaction strength and $L$ is the system length. The potential $V(x,t)$ models the optical lattice potential. Both
the repulsive contact interaction and the optical lattice potential are time-dependent in order to
consider the evolution in the absence of interaction and to represent the experimental sequence.

In order to numerically simulate this system, we discretize \cite{KollathZwerger2003} the Hamiltonian as
\begin{align}
  H &= -J \sum_{l=1}^{L_d} (b^\dagger_{l+1} b_l + \text{h.c.}) + \frac{U}{2} \sum_{l=1}^{L_d} n_l (n_l - 1) \nonumber \\
  &~~ + \sum_{l=1}^{L_d} V_l(t)~n_l, \label{eq:discretizedH}
\end{align}
where $J~\Delta x^2 = \hbar^2/(2m)$, $U \Delta x = g^\text{1D}$, $b_l$ and $b_l^\dagger$ are the annihilation
and creation operators for the atoms located at the discretized positions $x = l~\Delta x$ where $l = 1, ..., L_d$ with $L_d=L/\Delta x$.
We assume periodic boundary conditions.
 
\begin{figure}[h!]
\includegraphics[width=.9\columnwidth,clip=true]{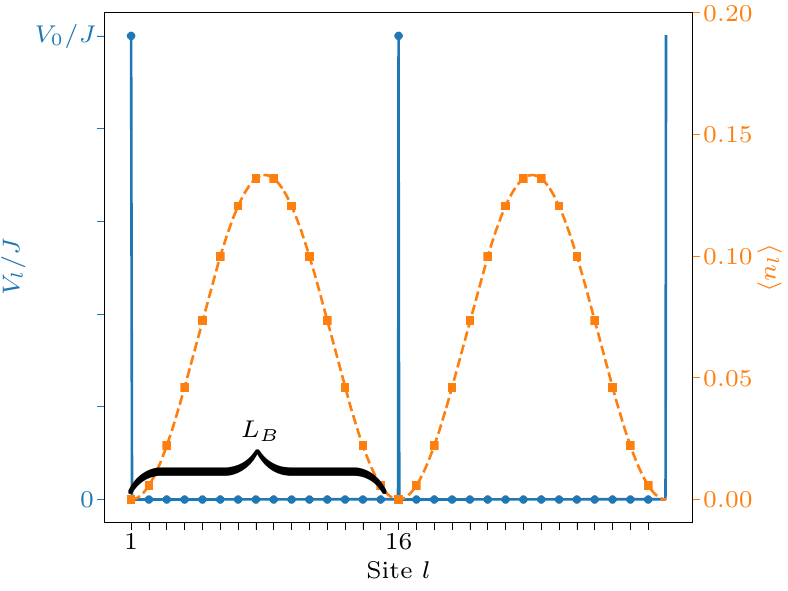}
\caption{Potential, $V_l$ (solid blue line), used to model the optical lattice: sharp barriers divide the systems into $N_B$
  wells. In the example shown here two wells, $N_B=2$, are discretized into $L_B=15$ sites each (including the
  site where the barrier is located). The initial density distribution, $\langle n_l \rangle$, obtained
  numerically for $U=0$ and $N=2$ (orange square markers) agrees well with a sinusodial profile (dashed orange line).
}
\label{fig:potential}
\end{figure}

\begin{figure}[h!]
\includegraphics[width=.9\columnwidth,clip=true]{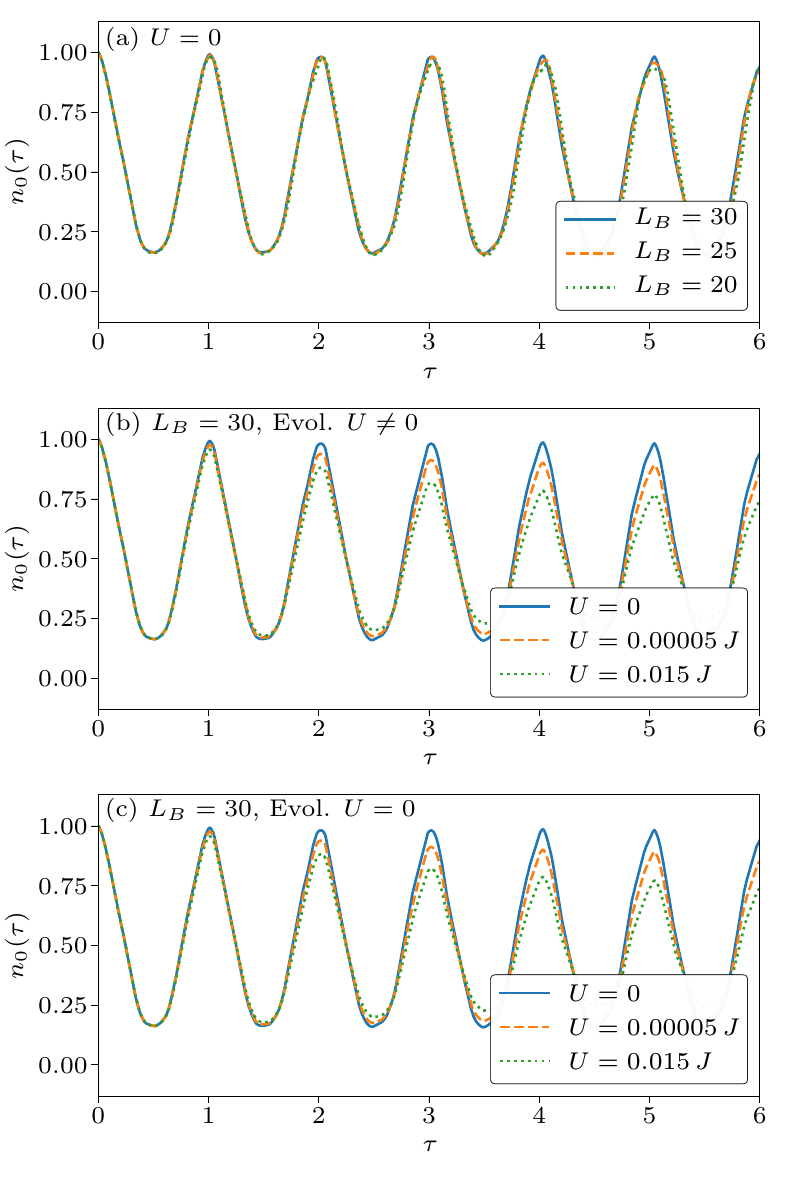}
\caption{Overlap $n_0$ as a function of normalized time, $\tau = t/T_{T}$, for 
  a system of $N_B = 20$ potential
    wells and $N = 2$ atoms. Panel (a): evolution in the absence of interaction, $U=0$,
    for three different discretizations $L_B = \{20, 25, 30\}$. The oscillations are weakly damped; however,
    this damping, not expected from an ideal non-interacting gas, is caused by the finite discretization and
    finite system effects. One can see here that the damping weakens with 
    increasing discretizations.
    Panel (b): evolution in the presence of interaction both during the system preparation and expansion
    (protocol (i)). We consider three different interaction strengths $U = \{0J, 0.00005J, 0.015J\}$ and $L_B = 30$. 
    The oscillation damping depends strongly on the interaction strength: larger interaction
    strengths cause a more rapid damping. Panel (c):  evolution in the absence of interaction during the expansion,
    interactions are only non-zero during the system preparation (protocol (ii)). The system is prepared using the same
    interaction strengths as in (b).
  }
\label{fig:timeevolution}
\end{figure}

The potential $V_l(t)$ is used to model the optical lattice potential which is switched off at
a certain moment in time. As numerically we are confined to the simulation of small systems, we use a
simplified potential consisting of very sharp $\delta$-peaks, i.e.~
\begin{align}
  V_l(t) = V_0 (1 - \Theta(t))\sum_{n=1}^{N_B} ~\delta_{l,(n-1) L_B + 1} 
\end{align}
where $N_B$ barriers of height $V_0$ are located at the discretized positions $x = [(n-1) L_B + 1] \Delta x$
where $L_B$ is the number of discretized sites within one potential well (including the barrier site on the left)
such that $L_B N_B = L/\Delta x$, and $\Theta(t)$ is the Heaviside function. Hence, $V_{l=1}(t) = V_{l=L/\Delta x + 1}(t)$.
In Fig.~\ref{fig:potential}, we provide a sketch of this potential for a system 
consisting of two
potential wells ($N_B = 2$) where each well is discretized into $L_B = 15$ sites. Due to the use of periodic boundary conditions,
one can notice that the third barrier on the right corresponds to the first barrier on the left. The initial density
distribution, $\langle n_l \rangle$, obtained numerically for a non-interacting system
containing two atoms, $N=2$, is also shown and agrees well with a sinusodial profile. 

A system containing $N$ atoms is therefore initialized in the ground state of the Hamiltonian $H$ (Eq.~\ref{eq:discretizedH}) at $t < 0$.
We then consider two different protocols. In the first protocol (i), the optical lattice potential is turned off abruptly at $t=0$ and the
system is let to evolve for a certain duration $\Delta t$; while, in the second protocol (ii), the same protocol is used but the
system is let to evolve in the absence of interaction. We denote the evolved wave function as $|\Psi(\Delta t)\rangle$.
In the experiment after the evolution time $\Delta t$, the optical lattice is projected on again, and the
total released energy is measured. This procedure has for consequence to measure the energy of the system in the presence of the
optical lattice, i.e.~
\begin{align}
\langle \Psi(\Delta t)|H| \Psi(\Delta t)\rangle &= \sum_n  E_n |\langle \Psi(\Delta t)|n\rangle|^2
\end{align}
where $|n\rangle$ are the eigenstates of the Hamiltonian in the presence of the optical lattice potential and of a finite
interaction strength (Eq.~\ref{eq:discretizedH}).
As the overlap is maximum with the ground
state $|0\rangle$, we determine theoretically the evolution of the overlap
\begin{align}
  n_0(t) = |\langle \Psi(\Delta t)|0\rangle|^2.
\end{align}

For the case of a non-interacting system, where $U=0$, using this experimental protocol, it was
demonstrated in Ref.~\cite{Santra2017} that each maximum and minimum in the Talbot signal, $n_0(t)$,
corresponds to a consecutive phase correlator.

The time-evolution of the Talbot signal, $n_0$, is shown in Fig.~\ref{fig:timeevolution}, where $n_0$ is plotted as
a function of $\tau = t/T_\text{T}$ for increasing interaction values and for the two protocols stated above
(evolution in the presence or absence of interaction). When the initial state is prepared in the absence of interaction, $U=0$,
the signal displays clear oscillations with maxima close to integer multiples of the Talbot time, $T_\text{T} = (L_B^2 \hbar)/(2\pi J)$,
and minima close to half-integer values.
The oscillations are weakly damped. This damping which is not expected from an ideal non-interacting gas is caused by the
finite discretization and finite system effects. In order to illustrate these effects, we show in Fig.~\ref{fig:timeevolution} (a)
three different values of $L_B$. Whereas only weak deviations between these three different values of $L_B$ are seen initially,
even for $U=0$, small differences start showing up for later maxima. One should note that for finer discretizations the
damping of the maxima decreases approaching the ideal undamped situation.
Furthermore, after a certain time, finite size effects come into play and can possibly affect
the overall structure of the oscillations.
%

The presence of interactions considerably changes the Talbot signal. Whereas the period of the oscillations appears to remain
constant within our simulation accuracy, the damping of the oscillations depends strongly on the interaction strength. A larger interaction
strength causes a more rapid damping of the oscillations. This is true for both protocols (i) and (ii).

\begin{figure}[h!]
\includegraphics[width=.9\columnwidth,clip=true]{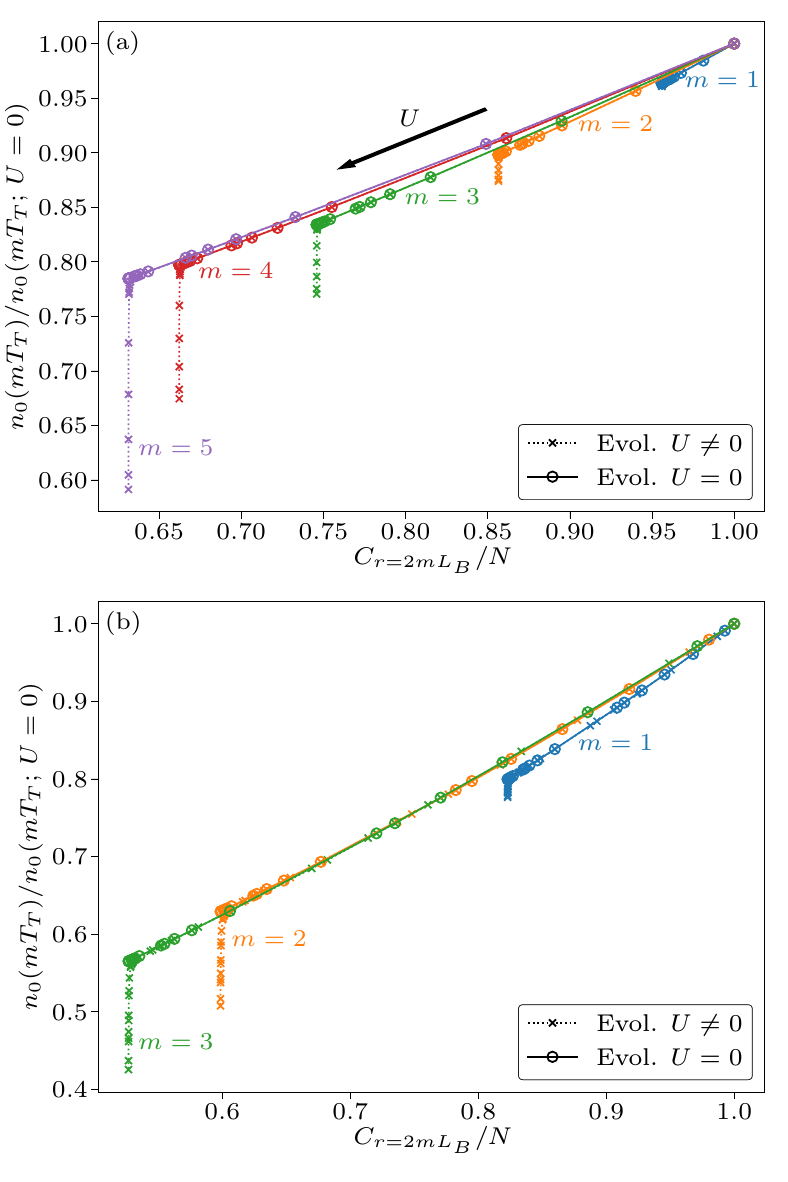}
  \caption{Maximum value of the overlap (occurring at $\tau = m T_{T}$ where $m$ is an integer) normalized by the corresponding
    overlap at $U=0$ as a function of the initial correlation $C_r$ at distance $r=2 m L_B $. The results for
    protocol (i) are marked with crosses whereare the ones for protocol (ii) are marked with circles. For both protocols, the interaction
    strength is varied in the range of $U=0$ up to $U= 0.5J$. Lower values of the initial correlations $C_r$
    correspond to larger values of the interaction strength $U$. Parameters are $L_B = 30$, with $N_B  = 20$ and $N = 2$ in panel (a)
    and $L_B=N_B = 15$, $N = 3$ in panel (b).}
\label{fig:relationN2andN3}
\end{figure}

We now turn our attention to the relationship between the maxima of $n_0$ and the single-particle correlations in
the initial state before the Talbot sequence is performed.
In the theoretical model, the initial single-particle
correlations between different lattice wells can be approximated as
\begin{align}
  C_r &= \sum_{l=1}^{L_d} \langle 0 | b^\dagger_{l+r} b_l | 0 \rangle 
\end{align}
where $r$ is a multiple of $L_B$ as here, due to the discretization, the spacing between consecutive sites of the optical lattice
is $d = L_B \Delta x$.

In the non-interacting situation, the value at integer and half integer
value of the Talbot time is directly related to the initial single-particle correlations at different distances~\cite{Santra2017}.
The height of the first maximum is proportional to the value of the correlation at distance $ r= 2 L_B$, i.e.~
\begin{equation}
\label{eq:relation}
C_{r = 2 m L_B} \propto n_0(t= m T_T),
\end{equation}
where $m = 1, 2, 3, \dots$ labels consecutive maxima.

In order to test whether the relation still holds in the presence of interaction, we plot the correlation $C_r$ versus the
value of the corresponding maxima in Fig.~\ref{fig:relationN2andN3} for two different particle
densities and both protocols (i) and (ii). We find that for low atom densities, the $m^\text{th}$ maximum, $n_0(m T_\text{T})$,
varies linearly with the $2m^\text{th}$ single-particle correlation, $C_{2 m L_B}$, as the interaction strength is weakly increased.
Lower initial values of the correlations correspond to larger interaction strengths. For weak interaction strengths, the proportionality
factor between $n_0(m T_\text{T})$ and $C_{2 m L_B}$ is approximately independent of $m$ and holds for both protocols.
Hence $n_0$ can be used in this interaction regime to measure the spatial single-particle correlations. However, 
at larger interaction strength, the relation breaks down for protocol (i). This breakdown is signalled
in Fig.~\ref{fig:relationN2andN3} by an abrupt downturn of the otherwise approximately linear curves
for large values of the interaction ($U \ge 0.025 J$).  The situation is different when the interaction
between the atoms is turned off at the same time as the optical lattice, i.e.~protocol (ii). In this case, the maxima of the
Talbot signal remain faithful representations of the single-particle correlations up to larger interaction strengths. This
indicates that the downturn is mainly due to the presence of interactions during the expansion process. Note, that for both protocols
the same values of the interaction strength are presented. 
Our results indicate that, even when the initial state is strongly interacting, the Talbot signal can
be used to measure the initial correlations as long as the interaction strength is lowered during the expansion.

A direct comparison of the critical interaction strength with the presented experimental results is not possible due to
the simplifying assumptions of the model. A truly quantitative comparison would require a more elaborate model, including
a realistic lattice potential and a fully three-dimensional treatment during the free expansion. However, the background
scattering length of the atoms in the current experiment is relatively weak and interaction effects in three dimension are
typically less effective than in one-dimensional setups. Therefore, the interpretation of the Talbot signal as a measurement
of the spatial single-particle correlations should remain intact for the presented experimental results.
Our theoretical findings also predict, that Talbot interferometers used
in molecular beam experiments, should provide confident measurements of the phase coherence in the matter wave field,
even in the presence of interactions, beyond single particle interference.

\section{Three-dimensional Talbot effect}
\label{sec:3d}

\begin{figure}
\includegraphics[width=0.9\columnwidth]{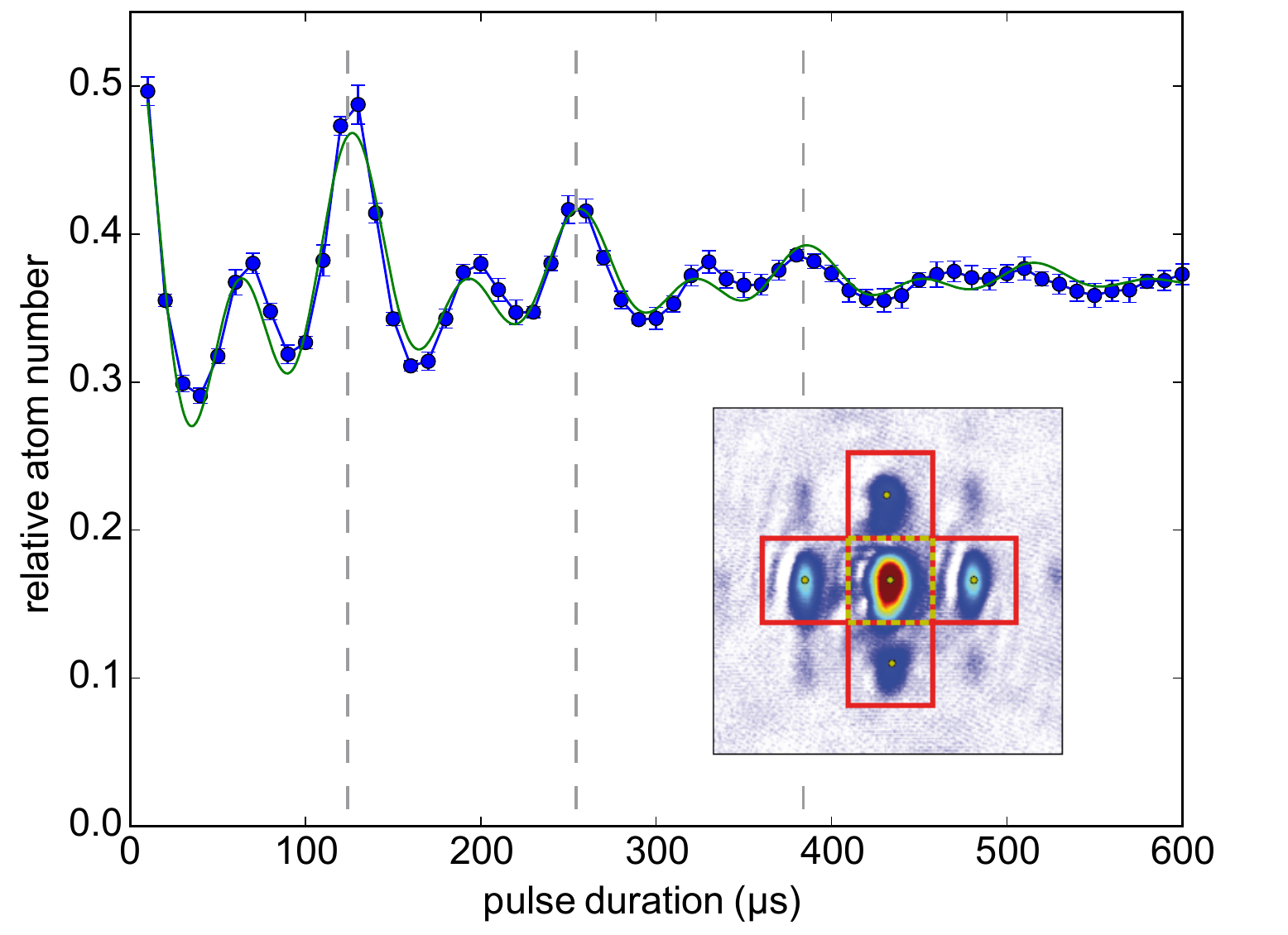}
\caption{Three-dimensional Talbot effect. The Talbot signal (relative atom number) is extracted from the time-of-flight absorption images (inset) by counting the number of atoms in the four first order diffraction peaks (boxes with solid line) normalized to the number of atoms in the central peak (box with dashed line). The two different Talbot times in the axial and transverse direction lead to a beat node in the Talbot signal. After $T_T=120\,\mu$s, the three-dimensional matter wave field is fully restored.}
\label{fig:3D_Talbot}
\end{figure}

While we restricted the discussion so far to a one-dimensional Talbot expansion, the presence of a three-dimensional optical lattice allows for a Talbot interferometer in all three spatial dimensions. As an initial state, we chose a superfluid in a lattice of the same lattice height as before ($s=5$). The Talbot interferometer now consists of the switching off and on of all three lattice axis simultaneously. During the free expansion the atoms can interfere in all three spatial dimensions. In the initial state, the interaction is identical to the previous experiment, while during the expansion, the interaction is lowered. Therefore, we expect that the interpretation as a measurement of the phase coherence remains intact. 

The transverse lattice directions have a lattice constant, which is a factor of $\sqrt{2}$ smaller than in the axial direction and the corresponding Talbot time is two times smaller, $T_\perp=60\,\mu$s. The interferometer signal is the beating between the two related frequencies and we expect full revivals after $T_T$ and partial revivals in between. Fig.\,\ref{fig:3D_Talbot} shows the result, displaying the expected behaviour. Thus, the full three-dimensional matter wave field is restored after $T_T$. 
 
\section{Conclusion}
We presented in this article results pertaining to 1D Talbot interference measurements in
a three-dimensional optical lattice in the presence of interactions. We theoretically validated the interpretation
of the Talbot signal as a measurement of spatial single-particle correlations of the intial state even 
when interactions are present during the state preparation. In contrast, the presence of strong interactions during the Talbot expansion
will render this relation invalid. Experimental results were shown to be in agreement with the expectations that
the amplitude of short range single-particle correlations decrease considerably with increasing lattice depths.
Additionally, a fully three-dimensional Talbot interferometer was created by switching off and on all lattice axis simultaneously.
Such an experiment would be very difficult to realize solely with light as it would require to spatially overlap
3D arrays of light sources and light detectors.

\section{Acknowledgments}
We would like to thank A. Pelster for discussions. We acknowledge financial 
support from the Deutsche Forschungsgemeinschaft (DFG, German Research Foundation) 
project number 277625399- TRR 185 project B3 and European Research Council (ERC) 
under the Horizon 2020 research and innovation programme, grant agreement No. 
648166 (Phonton). P.H. acknowledges financial support from the Bonn-Cologne 
Graduate School of Physics and Astronomy honors branch.
C.B. acknowledges financial support by DFG within the 
Graduate School of Excellence MAINZ, project number 49741853.

\bibliography{phoellmer_talbot_interaction}

\end{document}